# Physicists Attempt to Scale the Ivory Towers of Finance

*Physicists have recently begun doing research in finance, and even though this movement is less than five years old, interesting and useful contributions have already emerged. This article reviews these developments in four areas, including empirical statistical properties of prices, random-process models for price dynamics, agent-based modeling, and practical applications.*

During the past decade or so, many physicists have gone to Wall Street to work in finance. While the commonly heard statement that "Wall Street is the biggest single recruiter of physics PhDs" appears to be an urban legend, physicists working as *quants*—quantitative analysts—are now unquestionably common in large investment banks and other financial businesses.

More recently, a countermovement has emerged as physicists have begun writing research papers on finance and economics. While this work has yet to have a major impact on mainstream economics research, papers on finance are appearing with some frequency in physics journals, with a few publications in major science journals such as *Nature*, and occasionally even in economics journals. A new movement sometimes called *econophysics* has been established. Recently, about 200 people participated in the Third Annual Applications of Physics in Financial Analysis Conference (http://www.nbi.dk/APCA), held at Dublin's Trinity College in July 1999, where speakers included both practitioners and academics.

In the last five years, roughly 30 recent physics PhDs have addressed finance topics with their doctoral research. To paraphrase Jean-Phillipe Bouchaud, a pioneer in this area and the advisor of several such students: "Somebody has to train all the physics graduates going into banking and finance, and we want it to be us, not people from other disciplines. To do this we need to establish a scientific presence in the field." There is a widespread feeling among members of this movement that finance offers fertile terrain for physicists. It might be possible, for instance, to describe the aggregate behavior of financial agents using the tools of statistical physics. Combined with a fresh point of view, this might lead to some good science.

Not surprisingly, the few economists who have paid any heed at all view the entry of physicists into economics with considerable skepticism. Economics and finance, like physics, depend on a depth of domain-specific knowledge that takes years to master. Many physicists working in this area are poorly versed in the finance and economics literature. The point of view and problem-solving approach are quite different. The problems presented in modeling the physical and social worlds are not the same, and it is not obvious that methods that work well in physics will also work well in economics. With some justification, many economists think that the entry of physicists into their world reflects merely audac-

J. Doyne Farmer
*Santa Fe Institute*

ity, hubris, and arrogance. Physicists are not known for their humility, and some physicists have presented their work in a manner that plays into this stereotype. The cultural barrier between the two groups will be difficult to overcome.

This schism was already evident at a conference held at the Santa Fe Institute in 1988 titled "The Economy as an Evolving Complex System."[1] Roughly half the participants were economists and the other half physicists. Although many of the physicists were largely ignorant of economics, that did not prevent them from openly criticizing the economists. At one point, Nobel laureate Phil Anderson said, "You guys really believe that?" At another point, Larry Summers (now Secretary of the Treasury) accused physicists of having a "Tarzan complex." This was not just a turf war. Whether due to nature or nurture, this conference clearly showed that there is a deep epistemological divide between physicists and economists that is difficult to cross.

In this article, I am going to attempt a brief critical review of some of the work done in the past four or five years by physicists working on problems in finance. I will not discuss work that might be more broadly called economics, because I am too ignorant to do so. In the spirit of full disclosure, I should make my biases clear at the outset: My interest in finance stems from trading financial instruments at Prediction Company, where I was one of the founders, using directional forecasting models based on timeseries analysis of historical data. According to many mainstream economists, the highly statistically significant profits we made should have been impossible. My view of finance relies at least as much on conversations with traders as with academics. More importantly, my formative religious training was in physics. I am thus a highly biased reviewer. My only claim to impartiality is a wide exposure to fields outside of physics and a lack of involvement in the early stages of the econophysics movement; as a new entrant with many ideas of my own, my hope at the outset was that none of the juicy problems had been solved yet.

The topics presented at the Dublin econophysics conference included a variety of subjects, ranging from metaphorical models to empirically driven practical applications. I will single out a few highlights, dividing the presentations into four categories: empirical statistical regularities in prices, random-process models, agent-based models for price formation and market evolution, and practical applications, such as option pricing, risk control, and portfolio formation.

This article was explicitly commissioned to review work by physicists—a theme defined by cultural history and kinship relations rather than by the subject of scientific investigation. I write this review with some reluctance. I believe that disciplinary boundaries are dangerous and disciplines should be broken down or wholly eliminated. At the risk of spoiling the dramatic thread of this story, my conclusion in reviewing this work is that it indeed has value. That does not mean that I wish to argue that economists should move over and let physicists rule. Rather, I think that physicists have something to contribute, and I hope to encourage physicists and economists to work together.

## Empirical statistical regularities in prices

The distribution of price fluctuations is one of the most basic properties of markets. For some markets the historical data spans a century at a daily timescale, and for at least the last decade every transaction is recorded. Nonetheless, the price distribution's functional form is still a topic of active debate. Naively, central-limit theorem arguments suggest a Gaussian (normal) distribution. If $p(t)$ is the price at time $t$, the *log-return* $r_\tau(t)$ is defined as $r_\tau(t) = \log p(t + \tau) - \log p(t)$. Dividing $\tau$ into $N$ subintervals, the total log-return $r_\tau(t)$ is by definition the sum of the log-returns in each subinterval. If the price changes in each subinterval are independent and identically distributed (IID) with a well-defined second moment, under the central limit theorem the cumulative distribution function $f(r_\tau)$ should converge to a normal distribution for large $\tau$.

For real financial data, however, convergence is very slow. While the normal distribution provides a good approximation for the center of the distribution for large $\tau$, for smaller values of $\tau$—less than about a month—there are strong deviations from normality. This is surprising, given that the autocorrelation of log-returns is typically very close to zero for times longer than about 15 to 30 minutes.[2,3] What is the nature of these deviations from normality and what is their cause?

The actual distribution of log-returns has *fat tails*. That is, there is a higher probability for extreme values than for a normal distribution. As one symptom of this, the fourth moment is larger than expected for a Gaussian. We can measure this deviation in a scale-independent manner by using the *kurtosis* $k = \langle(r - \langle r \rangle)^4\rangle / \langle(r - \langle r \rangle)^2\rangle^2$ ($\langle \rangle$ in-dicates a time average). In the early 1960s, Benoit Mandelbrot[4] (now famous as the grandfather of fractals) and Eugene Fama[5] (now famous as the high priest of efficient market theory) presented empirical evidence that $f$ was a *stable Levy distribution*. The stable Levy distributions are a natural choice because they emerge from a generalization of the central limit theorem. For random variables that are so fat-tailed that their second moment doesn't exist, the normal central limit theorem no longer applies. Under certain conditions, however, the sum of $N$ such variables converges to a Levy distribution.[3] The Levy distributions are characterized by a parameter $1 \leq \mu \leq 2$, where $\mu = 2$ corresponds to the special case of a normal distribution. For $\mu < 2$, however, the stable Levy distributions are so fat-tailed that their standard deviation and all higher moments are infinite—that is, $\langle r^q \rangle = \infty$ for $q \geq 2$. In practice, this means that numerical estimates of any moment $q = 2$ or higher will not converge. Based on daily prices in different markets, Mandelbrot and Fama measured $\mu \approx 1.7$, a result that suggested that short-term price changes were indeed ill-behaved: if the variance doesn't exist, most statistical properties are ill defined.

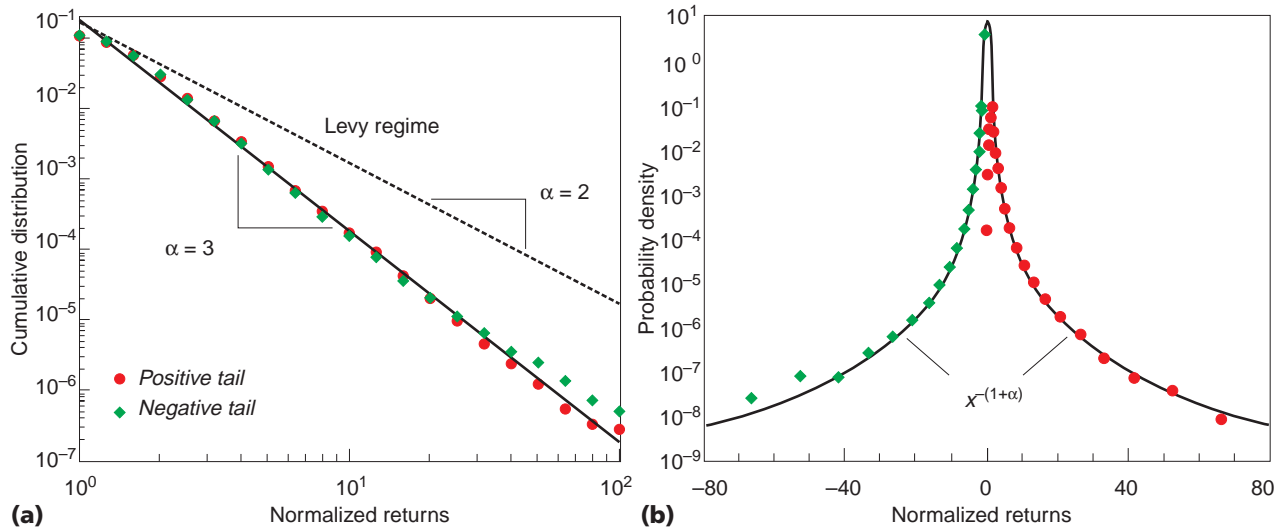

**Figure 1. Fat tails in price fluctuations: (a) Cumulative distribution of the positive and negative tails for normalized log-returns $r_\tau$ of 1,000 of the largest US companies for 1994–1995, with $\tau$ = 5 minutes.[12] The returns are normalized by dividing by the standard deviation for each company during this period. The solid line is a regression fit in the region $2 < r < 80$. (b) The probability density function of the normalized returns. The values in the center of the distribution arise from the discreteness in stock prices, which are important for small price movements.**

Subsequent studies demonstrated, however, that the behavior is more complicated than this.[6–12] First, for larger values of $\tau$, the distribution becomes progressively closer to normal. Second, investigations of larger data sets (including work by economists in the late '80s and early '90s[6–8]) make it clear that large returns asymptotically follow a power law $f(r) \sim |r|^{-\alpha}$, with $\alpha > 2$. This finding is incompatible with the Levy distribution. The difference in the value of $\alpha$ is very important: with $\alpha > 2$, the second moment (the variance) is well defined. A value $2 < \alpha < \infty$ is incompatible with the stable Levy distribution and indicates that simply generalizing the central limit theorem with long tails is not the correct explanation.

Physicists have contributed to this problem by studying really large data sets and looking at the scalings in close detail. A group at Olsen and Associates, led by Michel Dacorogna, studied intraday price movements in foreign exchange markets.[9] Another group at Boston University, led by Rosario Mantegna and Eugene Stanley, has studied the intraday movements of the S&P index.[10,11] More recently, they studied the five-minute returns of 1,000 individual stocks traded on the AMEX, NASDAQ, and NYSE exchanges, over a two-year period involving roughly 40 million records.[12] In this case, they observed the power-law scaling over about 90 standard deviations (see Figure 1). For larger values of $|r|$, these results dramatically illustrate that $f(r)$ is approximately a power law with $\alpha \approx 3$. Thus, the mean and variance are well-defined, the kurtosis clearly diverges, and the behavior of the skewness is not so clear.

Power-law scaling is not new to economics. The power-law distribution of wealth discovered by Vilfredo Pareto (1848–1923) in the 19th century predates any power laws in physics.[13] And indeed, since Pareto, the existence of power laws has been controversial. One underlying reason is that power-law probability distributions are necessarily approximations. An inverse power-law cumulative distribution $f(r) \sim |r|^{-\alpha}$ with an exponent $\alpha > 0$ is not integrable at zero, and similarly, with an exponent $\alpha \leq 0$, it is not integrable at infinity. Thus, a power-law probability distribution cannot be exactly true for a variable with an unbounded range. When they apply at all, power-law distributions are necessarily only part of a more complete description, valid within certain limits. (See the "Power law distribution of wealth" sidebar for more on this topic.[14])

Another reason for skepticism about power laws in economics is that sloppy statistical analysis has led to mistakes in the past. In the 1980s, there was considerable interest in the possibility that price changes might be described by a low-dimensional chaotic attractor. Physics and biology have many examples where the existence of low-dimensional chaos is unambiguous. Why not economics? Based on a numerical computation of fractal dimension, several researchers claimed to observe low-dimensional chaos in price series. Such computations are done by measuring the coarse-grained size of a set, in this case a possible attractor of returns in a state space whose variables are lagged returns, as a function of the scale of the coarse-graining. If this behaves as a power law in the limit where the scale is small, it implies low-dimensional chaos. But it is very easy to be fooled when performing such calculations. It is critical to test against a carefully formulated null hypothesis.[15] More careful statistical analysis by José Scheinkman and Blake LeBaron showed that the claims of low-dimensional chaos in price series were not well-justified.[16] While nonlinearity is clearly present, there is no convincing evidence of low-dimensionality. The power-law scaling that people thought they saw was apparently just an artifact of the finite size of their data sets.

The power law for large price moves is a very different story. To detect a chaotic attractor based on its fractal dimension in state space requires a test of the distribution's fine-grained, microscopic properties. Low-dimensional chaos is a very strong hypothesis, because it would imply deep structure and short-term predictability in prices. A power law in the tails of the returns, in contrast, is just a statement about the frequency of large events and is a much weaker hypothesis. This becomes clear in the context of extreme value theory. For simplicity, consider the positive tail $r \to \infty$. Under very general conditions, there are only three possible limiting behaviors, which we can classify based on the tail index $\alpha$:

1. There is a maximum value for the variable. The distribution vanishes for values greater than this maximum, and $\alpha < 0$.
2. The tails decay exponentially and $1/\alpha = 0$ (an example is a normal distribution).
3. There are fat tails that decay as a power law with $\alpha > 0$.

Price returns must be in one of these three categories, and the data clearly points to choice 3 with $\alpha > 2$.[2,6–12] Surprisingly, this implies that the price-formation process cannot be fully understood in terms of central limit theorem arguments, even in a generalized form. Power-law tails do obey a sort of partial central limit theorem: For a random variable with tail exponent $\alpha$, the sum of $N$ variables will also have the same tail exponent $\alpha$.[17] This does not mean that the full distribution is stable, however, because the distribution's central part, as well as the power law's cutoff, will generally vary. The fact that the

### Power-law distribution of wealth

Sorin Solomon and various collaborators have proposed an interesting idea for understanding the power-law distribution of wealth. An old approach to such an understanding entails modeling an agent's wealth at any given time as a multiplicative random process. By taking logarithms, it is clear that in the absence of any constraints such a process will tend toward log-normal distribution. When the mean is small compared to the standard deviation, an approximate power-law distribution for large wealth arises. The problem is that the tail exponent $\alpha = 1$, whereas the observed exponent varies, but can be above 2 or even 3. Solomon and his colleagues suggest that we can fix this problem by renormalizing for drifts in total wealth and imposing a constraint that there is a minimum relative wealth below which no one is allowed to drop. The exponent of the power law is then $\alpha = 1/(1 - c)$, where $c$ is the minimum wealth expressed as a fraction of the mean wealth. The Pareto exponent should thus be higher for countries that have a higher floor of minimum income, meaning a stronger welfare system.

distribution's shape changes with $\tau$ makes it clear that the random process underlying prices must have nontrivial temporal structure, as I'll discuss next. This complicates statistical analysis of prices, both for theoretical and practical purposes, and gives an important clue about the behavior of economic agents and the price-formation process. But unlike low-dimensional chaos, it does not imply that the direction of price movements is predictable. (Also see the "Power-law scaling" sidebar.[18])

**The search for a random process model of prices**

Price fluctuations are not identically distributed. Properties of the distribution, such as the variance, change in time. This is called *clustered volatility*. While the autocorrelation of log-returns, $\rho(\tau) \sim \langle r_\tau(t + \tau) r_\tau(t) \rangle$, is generally very small on timescales longer than a day, this is not true for the volatility (which can be defined, for example, as $r^2$ or $|r|$). The volatility on successive days is positively correlated, and these correlations remain positive for weeks or months. Clustered volatility can cause fat tails in $f(r)$. For example, the sum of normally distributed variables with different variances has a high kurtosis (although it does not have power-law tails).[3] To understand the statistical properties of price changes, we need a more sophisticated model that accounts for the probability distribution's temporal variation.

Clustered volatility is traditionally described by simple ad hoc time-series models with names that include the letters ARCH (for AutoRegressive Conditional Heteroscedasticity).[19,20] Such models involve linear relationships between the square or absolute value of current and past log-returns. Volatility at one time influences volatility at subsequent times. ARCH-type models can be effective for forecasting volatility, and there is a large body of work devoted to problems of parameter estimation, variations on the basic model, and so forth. ARCH models are not compatible with all of the empirical properties of price fluctuations, however.

A good price-fluctuations model should connect the behavior on multiple timescales. A natural test is the behavior of moments, in this case $\langle |r_\tau|^q \rangle$ as a function of $q$ and $\tau$. Several groups report approximate power-law scaling with $\tau$, with different slopes for each value of $q$, as Figure 2 shows.[21,22] In the jargon of dynamical systems theory, this suggests a fractal random process. A slope that is a linear function of $q$ implies a simple fractal process, and a slope that is a nonlinear function of $q$ implies a *multifractal* or *multiscaling* process. Indeed, several different calculations seem to show that the slope varies nonlinearly with $q$, suggesting that the price process is multifractal.

These results have suggested a possible analogy to turbulence.[21] Under this analogy, velocity plays the role of the logarithm of price and length plays the role of time. The hypothesis is that there is an analogy between the Kolmogorov energy cascade, through which large vortices break up into smaller vortices, and an *information cascade*, in which financial agents with more money or longer-term strategies influence financial agents betting smaller sums over shorter spans of time, inducing a cascade of volatility. This view is reinforced by Alain Arneodo, Jean-Francois Muzy, and Didier Sornette, who use a wavelet decomposition of volatility and an analysis in terms of mutual information to argue that there is indeed a cascade of volatility from large to small scales.[23]

While this is an exciting and interesting idea, caution is in order. As Mantegna and Stanley discussed, turbulence and price fluctuations differ in many important ways.[2] There is a possible alternative explanation for the complicated scaling behavior of price fluctuations.[24,3] Numerous studies show clearly that the autocorrelation function for volatility decays as a power law, $g(\tau) \sim \tau^{-\nu}$, with $\nu$ somewhere in the range $0.1 < \nu < 0.3$.[2,3,21–28] In this case, Jean-Phillipe Bouchaud, Marc Potters, and Martin Meyer show that the higher moments automatically scale as sums of power laws with different slopes. Asymptotically, the dominant power law has an exponent proportional to $q$, but for smaller values of $\tau$, another power law whose exponent is a nonlinear function of $q$ might dominate. Thus, there is apparent multifractal behav-

---

**Power-law scaling**

Stanley's group at Boston University has discovered a new power-law scaling that economists apparently had not previously observed. They show that the fluctuations in the growth rates of companies of size $S$ follow a power law $S^{-\beta}$, with $\beta \approx 0.2$ The growth-rate fluctuation is measured by the standard deviation of the distribution of a variety of different quantities, including sales, number of employees, assets, and several other measures. Furthermore, they observe a similar power law with approximately the same exponent for the GNPs of countries. This result is particularly interesting because it is not explained by standard theories (and they have proposed a new theory).

ior, even though asymptotically the process is just a simple fractal, with all the moments determined by the scaling of a single moment. For a short data set, a simple fractal process might look like a multifractal process due to slow convergence.

At this point, we can't say which of these two explanations is correct. Unlike fluid turbulence, where multifractal scaling is supported by strong theoretical evidence and by analysis of very large data sets, the situation in finance is still not clear. Resolution of this problem will await analysis of longer data sets and development of better theoretical models.

One thing that does seem clear is that conventional ARCH-type models are incompatible with the scaling properties of price fluctuations.[2,3] While ARCH-type models can indeed give rise to fat-tailed probability distributions with $\alpha > 2$, they cannot explain other properties of the price fluctuations.[28] ARCH-type models fit at a given timescale $\tau$ do not appear to do a good job of explaining the volatility at a different timescale $\tau$. Furthermore, conventional ARCH models do not have asymptotic power-law decay in the volatility autocorrelation function. The most likely explanation is that ARCH models are misspecified—their simple linear structure is not general enough to fully capture the real temporal structure of volatility. Given that they are completely ad hoc models, this is not surprising.

There are still missing pieces and several open questions to be answered before we will have a good random-process model linking the behavior of prices across a range of different time scales. Physicists have contributed to the theory and data analysis leading to the current understanding. They have also contributed an interesting new hypothesis; even if the analogy to turbulence turns out to be wrong, it has already stimulated interesting alternatives. But to have a good theory of how prices behave, we will need to explain the behavior of the agents on whom they depend. See the "Agent-based models" sidebar for a discussion of these models.

## Practical applications

Let's now look at some of the practical applications that have come out of the physics community. Physicists have taken the fat tails of the price distribution seriously and explored their implications in several areas, such as risk control, portfolio formation, and option pricing. In addition, they have made use of results on random matrices to provide some insight into the prob-lem of estimating correlation matrices between different assets.

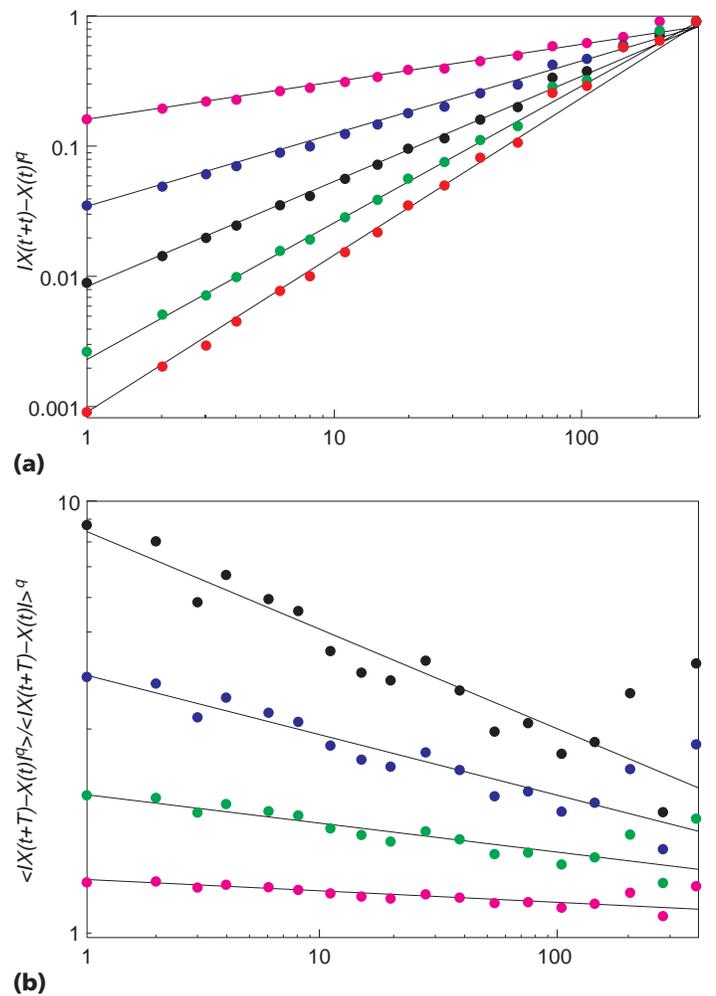

**Figure 2. Power-law scalings: (a) The mean moment of the absolute daily returns $\langle |r_t|^q \rangle$ plotted as a function of $\tau$ for several different values of $q$ ranging from 0.5 (top) to 2.5 (bottom).[22] In a log-log representation these appear to be approximately straight lines, suggesting power-law behavior. Furthermore, the variation of this line's slope with $q$ looks nonlinear, suggesting multifractal behavior. (b) As a test of this, the ratio $\langle |r|^q \rangle / \langle |r| \rangle^q$, which is constant for a simple fractal, but not for a multifractal for $q = 1.5$–3 (bottom to top).**

Perhaps the most direct consequence of fat tails is their impact on risk control. With fat tails, the probability of extreme events can be orders of magnitude larger than it is with a normal distribution. For a fat-tailed distribution, the variance is an inadequate and potentially misleading indicator of risk. Failure to take this into account can be disastrous. This was dramatically illustrated in October 1998 by the near failure of Long Term Capital Management, which was apparently at least in part due to a lack of respect for fat tails.

## Agent-based models

The elementary building blocks of financial markets are human agents, each buying and selling based on his or her own point of view. To have a full and fundamental description of markets requires models that take intelligence, learning, and complex decision making into account, not to mention emotion, irrationality, and intuition.

The traditional view in economics is that financial agents are completely rational with perfect foresight. Markets are always in equilibrium, which in economics means that trading always occurs at a price that conforms to everyone's expectations of the future. Markets are efficient, which means that there are no patterns in prices that can be forecast based on a given information set. The only possible changes in price are random, driven by unforecastable external information. Profits occur only by chance.

In recent years this view is eroding.[1] Modern economic theory assumes bounded rationality.[2] Equilibria are dynamic, driven by agents' changing perceptions of each others' actions. Allowance is made for the possibility of reasonable excess profits for agents who perform services, such as reducing risk or processing information. The *behavioral economists* have presented evidence of irrational behavior and market anomalies that historically would have allowed excess profits.[3] Anecdotal evidence suggests that some individuals might indeed make statistically significant excess profits.

It is fair to say that the physicists studying these problems tend toward the more radical end of the spectrum. While bounded rationality is a nice idea, it is only part of the story. People are not identical finite-capacity calculating machines differing only in their utility functions. Equally important is the diversity of viewpoints induced by nature and nurture. Formulating successful predictive models is extremely difficult and requires both hard work and intelligence. To make a good model, it is necessary to specialize, which stimulates diversification of financial strategies. As a result, financial agents are very heterogeneous. Some agents are more skilled than others, and the excess profits of such agents are not necessarily reasonable. The behavioral economists are clearly right that people are not fully rational and that this can play an important role in setting prices.[3,4] But where do we go from there? Despite the idiosyncrasies of human psychology, is there a *statistical mechanics* that can explain some of the statistical properties of the market, and perhaps take such idiocyncracies into account?

Agent-based modeling offers one approach to addressing these problems.[5] Efforts in this direction range from simple, metaphorical models, such as those of evolutionary game theory,[6] to complicated simulations, such as the Santa Fe Institute stock market model.[7,8] The SFI model, which was a collaboration between two economists, a physicist, and a computer scientist, was a significant accomplishment. It demonstrated that many of the dynamical properties of real markets, such as clustered volatility and fat tails, emerge automatically when a market simulation allows the views of the participants to be dynamic. It was a good start, but in part because of the complexity of the numerical simulations, it left many unanswered questions.

### The minority game

The minority game represents the opposite end of the spectrum. Despite its simplicity, it displays some rich behavior. While the connection to markets is only metaphorical, its behavior hints at the problems with the traditional views of efficiency and equilibrium. The minority game was originally motivated by Brian Arthur's El Farol problem.[9,10] El Farol is a bar in Santa Fe, near the original site of the Santa Fe Institute, which in the old days was a popular hangout for SFI denizens. In the El Farol problem, a fixed number of agents face the question of whether or not to attend the bar. If the bar is not crowded, as measured by a threshold on the total number of agents, an agent wins if he or she decides to attend. If the bar is too crowded, the agent wins by staying home. Agents make decisions based on the recent record of total attendance at the bar. This problem is like a market in that each agent tries to forecast the behavior of the aggregate and that no outcome makes everyone happy.

The minority game introduced by Damien Challet and Yi-Cheng Zhang is a more specific formulation of the El Farol problem.[11–15] At each timestep, $N$ agents choose between two possibilities (for example, A and B). A historical record is kept of the number of agents choosing A; because $N$ is fixed, this automatically determines the number who chose B. The only information made public is the most popular choice. A given time step is labeled "0" if choice A is more popular and "1" if choice B is more popular. The agents' strategies are lookup tables whose inputs are based on the binary historical record for the previous $m$ timesteps. Strategies can be constructed at random by simply assigning random outputs to each input (see Table A).

Each agent has $s$ possible strategies, and at any given time plays the strategy that has been most successful up until that point in time. The ability to test multiple strategies and use the best strategy provides a simple learning mechanism. This learning is somewhat effective—for example, asymptotically A is chosen 50% of the time. But because there is no choice that satisfies everyone—indeed, no choice that satisfies the majority of the participants—there is a limit to what learning can achieve for the group as a whole.

**Table A. Example of a strategy for the minority game. The input is based on the attendance record for the $m$ previous timesteps, 0 or 1, corresponding to which choice was most popular. In this case $m = 2$. The output of the strategy is its choice (0 or 1). Outputs are assigned at random.**

| Input | Output |
|-------|--------|
| 0 0   | 1      |
| 0 1   | 0      |
| 1 0   | 0      |
| 1 1   | 1      |

When $s > 1$, the sequence of 0s and 1s corresponding to the attendance record is aperiodic. This is driven by switching between strategies. The set of active strategies continues to change even though the total pool of strategies is fixed. For a given number of agents, for small $m$ the game is efficient, in that prediction is impossible, but when $m$ is large, this is no longer the case. In the limit $N \to \infty$, as $m$ increases there is a sharp transition between the efficient and the inefficient regime.

The standard deviation of the historical attendance record, $\sigma$, provides an interesting measure of the average utility. Assume that each agent satisfies his or her utility function by making the minority choice. The average utility is highest when the two choices are almost equally popular. For example, with 101 agents the maximum utility is achieved if 50 agents make one choice and 51 the other. However, it is impossible to achieve this state consistently. There are fluctuations around the optimal attendance level, lowering the average utility. As $m$ increases, $\sigma$ exhibits interesting behavior, starting out at a maximum, decreasing to a minimum, and then rising to obtain an asymptotic value in the limit as $m \to \infty$ (see Figure A). The minimum occurs at the transition between the efficient and inefficient regimes.

The distinction between the efficient and inefficient regimes arises from the change in the size of the pool of strategies present in the population, relative to the total number of possible strategies. The size of the pool of strategies is $sN$. The number of possible strategies is $2^{2^m}$, which grows extremely rapidly with $m$. For example, for $m = 2$ there are 16 possible strategies, for $m = 5$ there are roughly 4 billion, and for $m = 10$ there are more than $10^{300}$—far exceeding the number of elementary particles in the universe. In contrast, with $s = 2$ and $N = 100$, there are only 200 strategies actually present in the pool. For low $m$, when the space of strategies is well-covered, the conditional probability for a given transition is the same for all histories—there are no patterns of length $m$. But when $m$ is larger, so that the strategies are only sparsely filling the space of possibilities, patterns remain. We can interpret this as meaning that the market is efficient for small $m$ and inefficient for large $m$.

The El Farol problem and minority game is a simple game with no solution that can satisfy everyone. This is analogous to a market where not everyone profits on any given trade. Studies of the minority game suggest that the long-term behavior is aperiodic: the aggregate behavior continues to fluctuate. In contrast to the standard view in economics, such fluctuations occur even in the absence of any new external information.

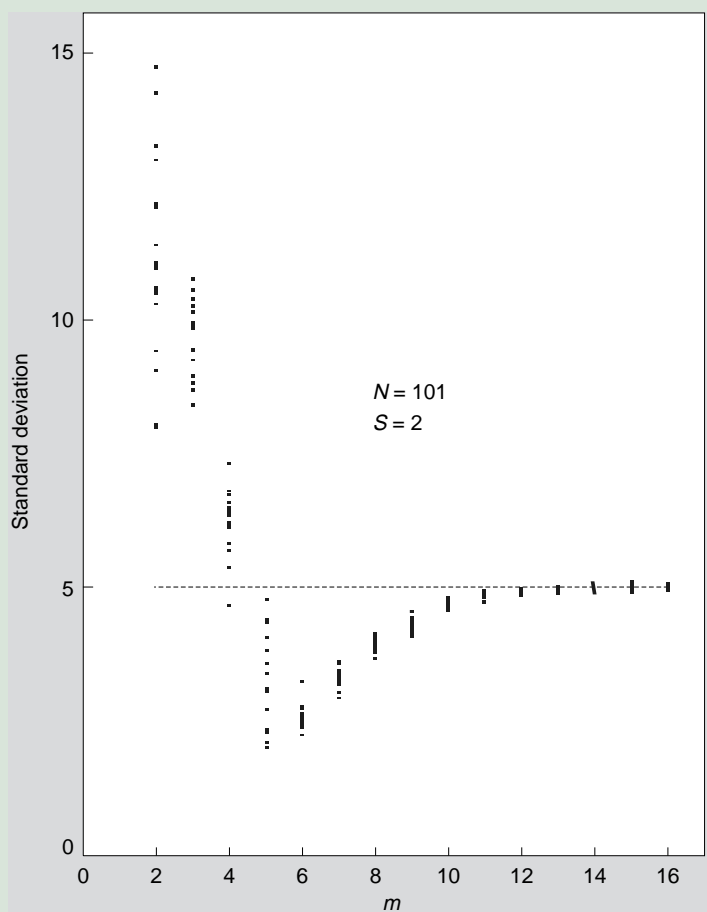

Figure A. For the minority game, the standard deviation of the attendance $\sigma$ is plotted as a function of the memory length of the strategies.[12] $\sigma$ is not zero, reflecting the fact that there are irregular oscillations in the attendance, with corresponding variations in the mean utility of the players of the game.

**Connections to financial markets**

While the results for the minority game are elegant and interesting in their own right, the connection to financial markets is only metaphorical. There are no prices and no trading in the minority game, and by definition a market is a place where prices are adjusted as trading takes place. Profits in markets are not made by being in the minority, but rather by anticipating the majority. To go past the metaphorical level, we must study models of markets that involve buying and selling assets and must be able to incrementally modify such models by adding successively more realism. There have been several steps in this direction.[16–26] These models involve a definition of agents, who make trading decisions, and a price-formation rule that determines how the price changes in response to these decisions. There is a feedback between these two processes: decisions affect prices, which affect decisions, and so forth. As a result, the prices and agents can display interesting dynamics.

Many dynamic trading models attempt to use strategies patterned after strategies actually used in real markets (as opposed to arbitrary, abstract strategies that have often been used in other game-theoretic models). The decisions are based on information internal to the market, such as prices and their history, and possibly information external to the market, and can be public (such as prices) or private (such as conversations between traders).

Despite the wide variation in financial-trading strategies, we can classify many of them into broad groups. Strategies that depend only on the price history are called *technical trading* or *chartist* strategies. The minority game strategies are technical-trading strategies (although not of a type that is widely used). *Trend-following strategies* are a commonly used special case of technical strategies in which the holdings of the asset positively correlate with past price movements. *Value* or *fundamental* strategies are strategies based on perceived value—that is, a model for what something ought to be worth, as opposed to its current price. The perceived value is inherently subjective, and from the point of view of the models discussed here is considered external information. Value strategies tend to buy when an asset is undervalued and sell when it is overvalued.

Many authors have studied trend and value strategies, with a variety of differences in the details of the implementations. Certain conclusions seem to emerge that are independent of the details. For example, trend strategies tend to induce trends and therefore positive autocorrelations in the price,[23] which was also evident in earlier work by economists.[27] Several new features, however, are apparent with a simple price formation rule that were not recognized in earlier studies by economists because of the more cumbersome framework they used to formulate the problem. For example, trend-following strategies also induce oscillations because, to keep risk bounded, strategies are formulated in terms of positions (holdings), whereas changes in price are caused by orders, which are changes in positions. Thus, the price dynamics have second-order, oscillatory terms.[23] Earlier studies also showed that trends are self-reinforcing—that trend strategies tend to induce trends in the market.[27] Some have mistaken this to mean "the more the merrier"—that the profits of trend strategies are enhanced by other identical trend strategies. A more careful analysis disproves this. While trend strategies indeed create profit opportunities, these opportunities are for *other* trend strategies, not for the *same* trend strategy.

A study of value-investing strategies also shows some interesting results.[23] Not surprisingly, most sensible value strategies induce negative autocorrelations in the price. Some value strategies cause prices and perceived values to track each other. But surprisingly, many sensible value strategies do not have this property. Another interesting set of questions concerns the case where the perceived values are heterogeneous (that is, people have different opinions about what something is worth). If all the strategies are linear, the market behaves just as though there were a single agent whose perceived value is the mean of the all the diverse values. If the strategies are nonlinear, however, the diversity of views results in *excess volatility*.[28]

All of the models I've discussed show certain generic phenomena, such as fat tails and clustered volatility, and preliminary results suggest that it might be possible to provide a quantitative explanation for some of the statistical properties of prices. For example, simulations by Thomas Lux and Michele Marchesi[21] (an economist and a physicist) use trend-following and value-investing strategies. They let agents switch from one group to the other. They observe power-law scaling in the tails of the log-returns, with a tail exponent $\alpha \approx$ 3, similar to that observed in real data. Other simulations also find power-law scaling in the tails in models that allow diffusion of private information.[26] For all of the models discussed in this sidebar, which allow dynamic interactions between prices and decisions, the problem is not, "How do we get realistic deviations from normality and IID behavior?" but rather, "How do we determine the necessary and sufficient conditions, and how do we identify which factors actually drive such effects in real markets?"

All of these models show variations in the price, reminiscent of boom-bust cycles. One source of these irregular cycles is the interaction between trend and value strategies. We can describe one such scenario roughly as follows: Suppose the market is strongly undervalued. The value investors buy, thereby raising the price. As the price rises, it creates a trend, causing trend followers to buy, raising the price even further. As the price becomes overvalued, the value investors sell, the trend is damped, and the trend followers eventually sell, and so on. In practice, there are many other effects and the resulting oscillations are highly irregular. See Figure B for an example illustrating the qualitative similarity of the resulting oscillations to those seen in real data. This example suggests that it would be interesting to try to develop more quantitative models of this type to see whether they are useful for prediction.

An economist would criticize the results in the studies I've cited for several reasons. A minor point concerns questions of whether the price-formation process some of these models use is sufficiently realistic. Use of an entirely ad hoc price-formation rule might cause inefficiencies or spurious dynamical effects. My results elsewhere have answered this criticism in part.[23] Perhaps more persuasively, all of these dynamic trading models have qualitative features such as clustered volatility and fat tails in common, even though the strategies and in some cases the price formation rules are quite different. So far, no careful studies have compared different methods of price formation or determined which market properties depend on the method of price formation. Perhaps the main contribution of physicists here has been the use of really simple methods for price formation, within which the dynamics are obvious and simple examples are easily solved analytically.

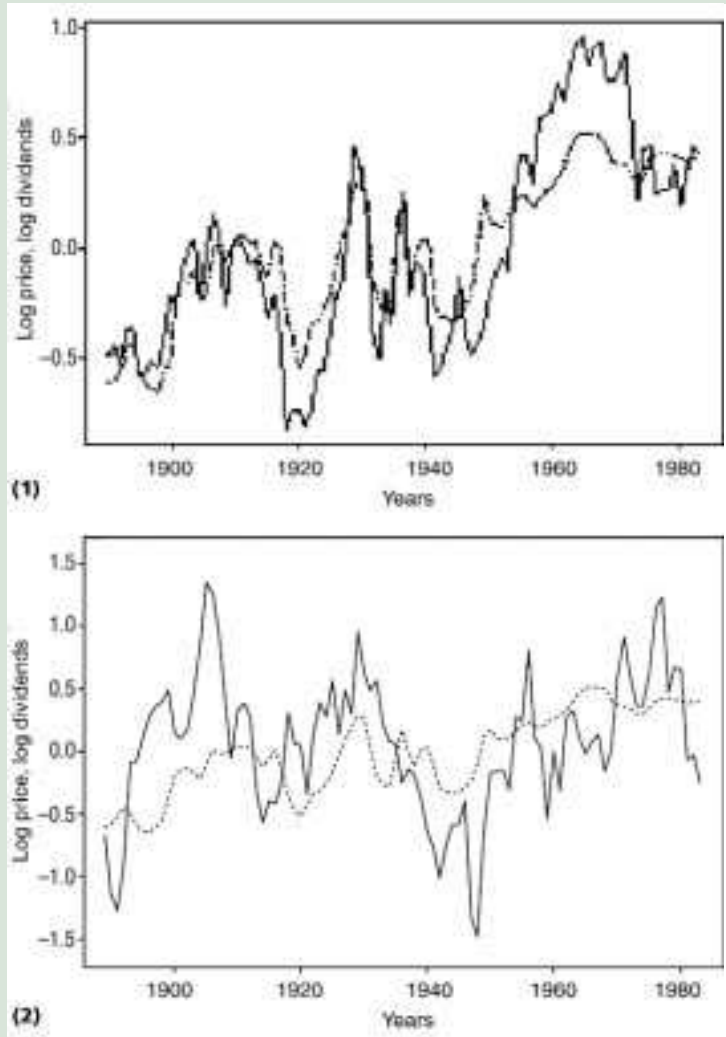

**Figure B. A comparison of the fluctuations in price and value of US stocks to an agent-based simulation:**[23] (a) the logarithm of the S&P index (solid), adjusted for inflation, compared to the logarithm of the mean dividend (dashed), which can be used as a crude measure of perceived value; (b) an agent-based simulation, using the same perceived value. The agents include a mixture of value and trend strategies. The relative population of the two groups is adjusted so that the net autocorrelation of the returns is zero, and the total population is adjusted to roughly match volatility. Otherwise there has been no attempt to adjust parameters or match initial states. The point is purely qualitative: For both the real data and the simulation, the price and the perceived value undergo large oscillations around each other, as the market becomes alternately underpriced and overpriced.

As a stronger criticism, many of these models contain only limited mechanisms for individual agents to learn, or perhaps more important, selection mechanisms that allow the agent population as a whole to learn. With appropriate selection mechanisms under standard dogma, the market should become efficient and the price should be random (or at least random enough that it is impossible for any agent to make profits).

Shareen Joshi and I have partially addressed this criticism.[24] We use a simple price-formation rule and employ the same representation of strategies discussed earlier for the minority game. For $s > 1$, the sequence of prices is aperiodic. We also add the ability for agents to reinvest their profits and accumulate capital, which provides a mechanism driving the market toward efficiency. Starting from an arbitrary initial state, the agents with profitable strategies accumulate capital. As this happens, their market impact increases, driving their profits down. The system approaches an attractor that in some respects resembles a classic economic equilibrium. Prices fluctuate irregularly.

Although asymptotically the capital of each agent ceases to increase, fluctuating around a mean value, before this state is reached agents with superior strategies accumulate capital. Unlike a classic equilibrium, the fluctuations are not driven by external information—the fluctuations are generated completely internally. Furthermore, the efficiency is only partial: for a reasonably large value of the memory, the entropy of the prices never approaches its maximum value. There are always patterns remaining in prices, representing profit-making opportunities for new strategies.

This study of dynamic agent-based trading models is still in its infancy, and many interesting problems remain to be addressed. To make these models more convincing. more work is needed to make them more realistic and better grounded in economic theory.

**References**

1. J.D. Farmer and A.W. Lo, "Frontiers of Finance: Evolution and Efficient Markets," *Proc. Nat'l Academy of Science,* Vol. 96, NAS, Washington, D.C., 1999, pp. 9991–9992.

The growing awareness of fat tails is changing the way people characterize risk. Ten years ago, sophisticated quantitative trading firms characterized risk in terms of standard deviations. Increasingly, this is changing to Value at Risk (VaR), the size of the loss that would be experienced with an unfavorable move of a given probability.[3] The probability level chosen is generally rather small—for example, 1%, where the normal begins to be a poor approximation. With a good estimate of the probability distribution of the returns on all the assets, we can use Monte Carlo methods to make a VaR estimate. This can be time-consuming, however. Jean-Phillipe Bouchaud and Marc Potters have recently offered an alternative.[29]

This method uses the optimal fluctuation method from physics to simplify VaR calculations by taking advantage of the fat tails. They expand the VaR in a Taylor series in terms of the derivatives of the value of a portfolio with respect to factors (such as principal values) and show that when the tails decay as a power law, higher derivatives can be neglected and a few factors usually dominate. This simplifies the VaR calculation, but more importantly, gives a better understanding of what risks depend on and how different risk factors interact.

Dealing with fat tails properly can also be important for constructing portfolios. The classic Markowitz approach is to maximize returns subject to a constraint on the variance of the portfolio. Although portfolio theory is often cited as one of the great achievements of modern finance theory, mean-variance portfolios tend to perform poorly in practice. Even if we assume that the underlying assets have a normal distribution, the portfolio weights and portfolio performance are so sensitive to estimation errors that the amount of data required to get good estimates can be prohibitive. This problem is much worse in the presence of fat tails.

Didier Sornette, Propsero Simonetti, and Jorgen Anderson recently introduced a new approach to portfolio formation that explicitly takes fat tails into account.[30] They make a change of variables into normally distributed coordinates using a simple numerical procedure and use the correlation matrix in these coordinates to form the portfolio. The resulting procedure is stabler and better conditioned than the usual mean-variance approach. They show that minimizing variance often increases large risks, as measured for example by VaR. With fat tails, an adequate prediction of the risks relies much more on a correct description of the tail structure than on the correlations between the assets.

The implications of fat tails are important for option pricing. An option is a financial instrument that gives the holder the option to buy or sell at a given price (the *strike price*) at a later time (the *expiration date)*. The value of an option depends on the strike price and expiration date, as well as on the underlying asset's statistical properties. The standard method of pricing options using the Black-Scholes formula assumes that the log-returns of the underlying asset are normally distributed. Under the Black-Scholes formula, we can compute the price of the option for a given volatility, or alternatively, the formula can be inverted to compute the implied volatility based on the option price.

According to the Black-Scholes formula, the

2. T.J. Sargent, *Bounded Rationality in Economics,* Clarendon Press, Oxford, UK, 1993.
3. W.F.M. deBondt and R.H. Thaler, "Financial Decision Making in Markers and Firms: A Behavioral Perspective," *Handbooks in Operations Research and Management Science*, Vol. 9, Elsevier Science, New York, 1995.
4. R. Shiller, "Human Behavior and the Efficiency of the Financial System," *Handbooks in Macroeconomics, Vol. 2*, Elsevier Science, New York, 1999.
5. B. LeBaron, "Agent-Based Computational Finance: Suggested Readings and Early Research," to appear in *J. Economic Dynamics and Control,* 1999.
6. J.W. Weibull, *Evolutionary Game Theory,* MIT Press, Cambridge, Mass., 1996.
7. W.B. Arthur et al., "Asset Pricing under Endogenous Expectations in an Artificial Stock Market," *The Economy as an Evolving, Complex System II*, Addison-Wesley, Reading, Mass., 1997.
8. B. LeBaron, W.B. Arthur, and R. Palmer, "Time Series Properties of an Artificial Stock Market," *J. Economic Dynamics and Control,* Vol. 23, 1999, pp. 1487–1516.
9. W.B. Arthur, "Inductive Reasoning and Bounded Rationality," *Am. Economic Assoc. Papers and Proc.*, Vol. 84, 1994, 406–411.
10. W.B. Arthur, "Complexity and the Economy," *Science,* Vol. 238, 1999, pp. 107–109.
11. D. Challet and Y.-C. Zhang, "Emergence of Cooperation and Organization in an Evolutionary Game," *Physica A,* Vol. 246, 1997, p. 407.
12. R. Savit, R. Manuca, and R. Riolo, "Adaptive Competition, Market Efficiency, Phase Transitions, and Spin-Glasses," *Physical Review Letters,* Vol. 82, 1999, p. 2203.
13. N.F. Johnson, M. Hart, and P.M. Hui, "Crowd Effects and Volatility in a Competitive Market," 1998; http://xxx.lanl.gov/cond-mat/ 9811227.
14. D. Challet , M. Marsili, and R. Zecchina, "Theory of Minority Games," 1999; preprint http://xxx.lanl.gov/cond-mat/99014392.
15. A. Cavagna et al., "A Thermal Model for Adaptive Competition in a Market," 1999;

implied volatility should be independent of the strike price. In practice, however, the implied volatility depends strongly on the strike price. For many assets, such as stocks, options with strike prices that are in the money (near the current price) have lower implied volatilities than those with strike prices that are out of the money. The implied volatility plotted against the strike price looks like a noisy parabola called the *smile*. The smile makes it clear that real option prices deviate from the Black-Scholes formula.

The Black-Scholes pricing theory has two remarkable features:

- the hedging strategies eliminate risk entirely, and
- the option price does not depend on the average return of the underlying asset.

There are very special properties that are only true under the assumption of normality. With a more realistic distribution for the underlying returns, risk in option trading cannot be eliminated and the correct option price depends on the full distribution of the underlying asset, including its mean. Physicists have played a leading role in developing practical and simple risk-return methods for pricing options in a more general context that takes into account deviations from normality, as well as transaction costs.[31,32] These are based on the principle that the proper option price minimizes (but does not eliminate) risk. In fact, in practice the residual risk is rather large, certainly much larger than the zero risk predicted by the Black-Scholes procedure. This results in good predictions of the smile. The "implied kurtosis" from such procedures also agrees reasonably well with the historical kurtosis. Furthermore, this procedure does a good job of predicting the dependence of the option price on the mean return, which is important to many practitioners, who might have views about the mean return. Science and Finance, a company consisting largely of physicists, has developed software employing these principles, which a major bank known for their expertise in pricing options is using.

Another interesting application in a somewhat different direction concerns the computation of correlation matrices. Correlation matrices are important for many reasons. Correlations are generally very important for hedging risks, for example, to optimize a portfolio using the conventional Markowitz approach. The correlation matrix for $N$ assets has $(N(N-1))/2$ independent values. Thus the computation of a correlation matrix is poorly determined for any large value of $N$ unless the effective length $T$ of the data sets is enormous. For example, to estimate the daily correlations of the stocks in the S&P index, just to have the number of data points equal the number of free parameters would require about 500 years of stationary data. The S&P index has not existed that long, the composition of companies is constantly changing, and the nature of companies changes, so that the relevance of price history in the distant past is questionable. Five to 10 years is typically the largest sensible value of $T$ for most practical applications. Estimation errors are a big problem.

To understand the structure of correlation matrices in such a highly random setting, physicists have recently applied the theory of random ma-

http://xxx.lanl.gov/cond-mat/9903415.

16. A. Rieck, "Evolutionary Simulation of Asset Trading Strategies," *Many-Agent Simulation and Artificial Life*, E. Hillebrand and J. Stender, eds., IOS Press, Amsterdam, 1994, pp. 112–136,

17. M. Levy, N. Persky, and S. Solomon, "The Complex Dynamics of a Simple Stock Market Model," *Int'l J. High Speed Computing*, Vol. 8, 1996, p. 93.

18. G. Caldarelli, M. Marsili, and Y.C. Zhang, "A Prototype Model of Stock Exchange," *Europhysics Letters*, Vol. 40, 1997, p. 479.

19. M. Youssefmir, B.A. Huberman, and T. Hogg, "Bubbles and Market Crashes," *Computational Economics*, Vol. 12, 1998, pp. 97–114.

20. P. Bak, M. Paczuski, and M. Shubik, "Price Variations in a Stock Market with Many Agents," *Physica A*, Vol. 246, 1997, pp. 430–453.

21. T. Lux and M. Marchesi, "Scaling and Criticality in a Stochastic Multiagent Model of Financial Market," *Nature,* Vol. 397, 1999, pp. 498–500.

22. J.-.P. Bouchaud and R. Cont, "A Langevin Approach to Stock Market Fluctuations and Crashes," *European Physics J. B*, Vol. 6, 1998, pp. 543–550.

23. J.D. Farmer, *Market Force, Ecology, and Evolution,* Santa Fe Inst. Working Paper 98-12-117, SFI, Santa Fe, N.M., 1998; http://xxx.lanl.gov/adapt-org 9812005.

24. J.D. Farmer and S. Joshi, *Evolution and Efficiency in a Simple Technical Trading Model*, SFI Working Paper 99-10-071, Santa Fe Inst., Santa Fe, NM, 1999.

25. D. Stauffer and D. Sornette, "Self-Organized Percolation Model for Stock Market Fluctuations," 1999; http://xxx.lanl.gov/cond-mat/9906434.

26. G. Iori, "A Microsimulation of Traders Activity in the Stock Market: The Role of Heterogeneity, Agents' Interactions, and Trade Frictions," 1999; http://xxx.lanl.gov/adp-org/0005005.

27. J.B. Delong et al., "Positive Feedback and Destabilizing Rational Speculation," *J. Finance*, Vol. 45, 1990, pp. 379–395.

28. R. Shiller, *Market Volatility,* MIT Press, Cambridge, Mass., 1989.

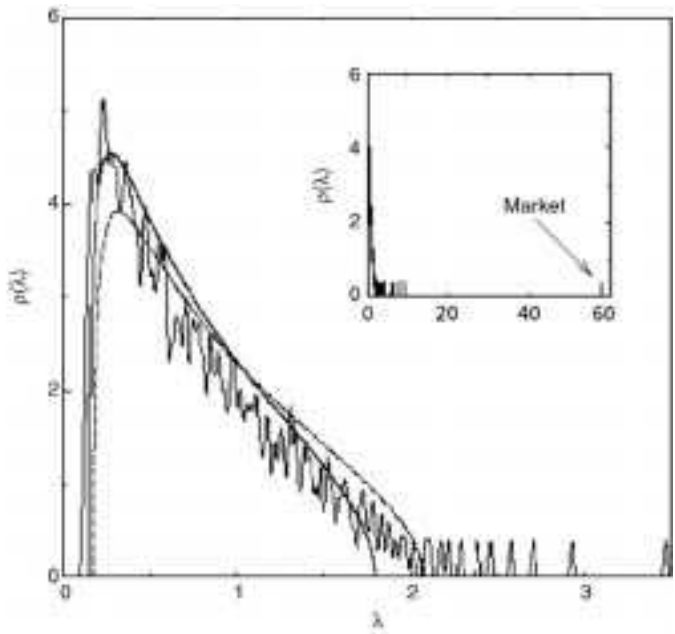

Figure 3. Smoothed density of the eigenvalues λ of an empirical correlation matrix for the returns of 406 companies in the S&P index, for the period 1991–1996.[33] For comparison, this is plotted against the theoretical density of eigenvalues if the matrix is completely random except for its largest eigenvalue (dotted line). A better fit is obtained by fitting parameters, as shown in the solid line. The inset shows the same thing when the largest eigenvalue, which corresponds to the overall market movement, is included.

trices, extensively developed in nuclear physics and elsewhere.[33,34] The eigenvalues and eigenvectors of random matrices approach a well-defined functional form in the limit $N \to \infty$. It is then possible to compare the distribution of empirically determined eigenvalues to the distribution that would be expected if the data were completely random, as shown in Figure 3. For the correlation matrix of 406 companies in the S&P index, in a computation based on daily data from 1991 to 1996, only seven out of 406 eigenvalues are clearly significant with respect to a random null hypothesis. This suggests that we can improve estimates by setting the insignificant eigenvalues to zero, mimicking a common noise-reduction method used in signal processing.

I have omitted quite a lot from this short review. Some of this is good work that is either out of the few main themes developed here or that is too complicated to explain with a small amount of space. Some is work that is either of poor quality or simply crazy, as is typical of new fields. It also reflects a difference between the intellectual cultures of physics and economics. Physics tends to be less restrictive about publication. In physics, publications appear more rapidly and peer review is typically not as strict. There is more emphasis on creativity and less on rigor. This is perhaps one of the luxuries of natural science, where theory is easily tested against data. Since their inception, it seems that mystics, cabalists, and alchemists have been attracted to the financial markets, and there are many such people who, knowing some mathematics, will always hover around any group that will pay them the courtesy of listening. In any case, a field should be judged by its best work, rather than its worst work.

At the Dublin conference, I was initially disturbed that almost all the speakers were physicists. This was apparently intentional. The conference's organizers felt that during the formative stages it is important not be too critical. Let new ideas grow and see where they lead; thinning can take place later. In the early stages, too much harsh criticism from economists might be counterproductive. Given the conservatism of many economists, there is some good sense to this. But the time is rapidly approaching when physicists who want to do serious work in finance need to interact more closely with economists. There were many bright people in Dublin and many good ideas. There was also sometimes a lack of grounding. Many of the physicists knew very few empirical facts about markets and were largely ignorant of the literature in economics and finance. Some of the work there focused on problems that might be fun from a mathematical point of view, but are not very relevant to understanding financial markets.

There are many fresh ideas in the work of physicists that have been missing or underrepresented in economics. These will lead to new and interesting work, and I expect rapid progress in the next few years. This will come as physicists do the hard work to master the domain knowledge. Physicists like me need to fully understand what has already been done by economists and stop reinventing the wheel. There have already been several good collaborations between physicists and economists, and hopefully there will be many more. While there is a hard core of conservative economists who will never accept these new ideas, they will eventually die. The beauty of the scientific method is that ultimately it is always possible for new ideas to gain acceptance if they are right. Prediction of empirical data based on elegant, concise scientific theories will ultimately triumph over dogma and myopia. There is no shortage of empirical data to test theories in finance and economics. The challenge for physicists is to understand the real problems, and produce theories that fit the data. A good start has been made, but the real challenges and adventures lie ahead.

## Acknowledgement

*I would like to thank Erik Aurell, Jean-Philippe Bouchaud, Michel Dacorogna, Blake LeBaron, David Sherrington, Didier Sornette, Francois Schmitt, and Eugene Stanley for valuable comments, and Andrew Lo for commissioning this article.*

## References

1. P.W. Anderson, J.K. Arrow, and D. Pines, eds., *The Economy as an Evolving Complex System,* Addison-Wesley, Redwood City, Calif., 1988.
2. R.N. Mantegna and H.E. Stanley, *Introduction to Econophysics: Correlations and Complexity in Finance*, Cambridge Univ. Press, Cambridge, UK, 1999.
3. J.-P. Bouchaud and M. Potters, *Theory of Financial Risk: From Statistical Physics to Risk Management,* Cambridge Univ. Press, Cambridge, UK, http://www.science-finance.fr.
4. B.B. Mandelbrot, "The Variation of Certain Speculative Prices," *J. Business,* Vol. 36, 1963, pp. 394–419.
5. E.F. Fama, "The Behavior of Stock Market Prices," *J. Business,* Vol. 38, 1965, pp. 34–105.
6. V. Akgiray, G.G. Booth, and O. Loistl, "Stable Laws Are Inappropriate for Describing German Stock Returns," *Allegemeines Statistisches Archiv.*, Vol. 73, 1989, pp. 115–121.
7. K.G. Koedijk, M.M.A. Schafgans, and C.G. De Vries, "The Tail Index of Exchange Rate Returns," *J. Int'l Economics*, Vol. 29, 1990, pp. 93–108.
8. T. Lux, "The Stable Paretian Hypothesis and the Frequency of Large Returns: An Examination of Major German Stocks," *Applied Financial Economics,* Vol. 6, 1996, pp. 463–475.
9. U.A. Müller, M.M. Dacorogna, and O.V. Pictet, "Heavy Tails in High-Frequency Financial Data," *A Practical Guide to Heavy Tails,* R.J. Adler, R.E. Feldman, and M.S. Taqqu, eds., Birkhäuser, Boston, 1998, pp. 382–311.
10. R.N. Mantegna and H.E. Stanley, "Scaling Behavior in the Dynamics of an Economic Index," *Nature,* Vol. 376, 1995, pp. 46–49.
11. P. Gopikrishnan et al., "Scaling of the Distribution of Fluctuations of Financial Market Indices," 1999; http://xxx.lanl.gov/cond-mat/ 9905305.
12. V. Plerou et al., "Scaling of the Distribution of Price Fluctuations of Individual Companies," 1999; http://xxx.lanl.gov/cond-mat/ 9907161.
13. J. Eatwell, M. Milgate, and P. Newman, *The New Palgrave: A Dictionary of Economics 3,* MacMillan Press, London, 1991.
14. O. Malcai, O. Biham, and S. Solomon, "Power-Law Distributions and Levy-Stable Intermittent Fluctuations in Stochastic Systems of Many Autocatalytic Elements," *Physical Rev. E*, Vol. 60, No. 2, 1998, pp. 1299–1303.
15. J. Theiler et al., "Testing for Nonlinearity in Time Series: The Method of Surrogate Data," *Physica D*, Vol. 58, 1992, pp. 77–94.
16. J. Scheinkman and B. LeBaron, "Nonlinear Dynamics and Stock Returns," *J. Business,* Vol. 62, 1989, pp. 311–338.
17. W. Feller, *An Introduction to Probability Theory and Its Applications,* Vol. 2, Wiley & Sons, New York, 1971.
18. Y. Lee at al., "Universal Features in the Growing Dynamics of Complex Organizations," *Physical Rev. Letters*, Vol. 81, No. 15, 1998, pp. 3275–3278.
19. R.F. Engle, "Autoregressive Conditional Hetereoskedasticity with Estimations of the Variance of UK Inflation," *Econometrica*, Vol. 50, 1982, pp. 987–1002.
20. J.Y. Campbell, A.W. Lo, and A.C. MacKinlay, *The Econometrics of Financial Markets*, Princeton Univ. Press, Princeton, N.J., 1997.
21. S. Ghashghaie et al., "Turbulent Cascades in Foreign Exchange Markets," *Nature*, Vol. 381 1996, pp. 767–770.
22. F. Schmitt, D. Schertzer, and S. Lovejoy, "Multifractal Analysis of Foreign Exchange Data," *Applied Stochastic Models and Data Analysis*, Vol. 15, 1999, pp. 29–53.
23. A. Arneodo, J.-F. Muzy, and D. Sornette, "'Direct' Causal Cascade in the Stock Market," *European Physical J. B,* Vol. 2, 1998, pp. 277–282.
24. J.-P. Bouchaud, M. Potters, and M. Meyer, "Apparent Multifractality in Financial Time Series," 1999; http://xxx.lanl.gov/cond-mat/ 9906347.
25. U.A. Müller et al., "Statistical Study of Foreign Exchange Rates, Empirical Evidence of a Price Scaling Law, and Intraday Analysis," *J. Banking and Finance*, Vol. 14, 1990, pp. 1189–1208
26. M.M. Dacaragona et al., "A Geographical Model for the Daily and Weekly Seasonal Volatility in the Foreign Exchange Market," *J. Int'l Money and Finance*, Vol. 12, 1993, pp. 413–428.
27. Z. Ding, C.W.J. Granger, and R.F. Engle, "A Long Memory Property of Stock Returns and a New Model," *J. Empirical Finance*, Vol. 1, 1993, p. 83.
28. P. Embrechts, C. Kluppelberg, and T. Mikosch, *Modelling Extremal Events*, Springer-Verlag, Berlin, 1997.
29. J.-P. Bouchaud and M. Potters, "Worst Fluctuation Method for Fat Value-at-Risk Estimates," 1999; http://xxx.lanl.gov/ cond-mat/ 9909245.
30. D. Sornette, P. Simonetti, and J.V. Andersen, "Field Theory for Portfolio Optimization: "Fat Tails" and Nonlinear Correlations," 1999; http://xxx.lanl.gov/cond-mat/9903203.
31. J.-P. Bouchaud and D. Sornette, "The Black-Scholes Option Pricing Problem in Mathematical Finance: Generalization and Extensions for a Large Class of Stochastic Processes," *J. Physics*, Vol. 4, 1994, pp. 863–881.
32. J.-P. Bouchaud, G. Iori, and D. Sornette, "Real-World Options: Smile and Residual Risk," *Risk 9*, Mar. 1996, pp. 61–65.
33. L. Laloux et al., "Noise Dressing of Financial Correlation Matrices," 1998; http://xxx.lanl.gov/cond-mat/9810255.
34. V. Plerou et al., "Universal and Nonuniversal Properties of Cross-Correlations in Financial Time Series," 1999: http://xxx.lanl.gov/cond-mat/9902283.

**J. Doyne Farmer** is the McKinsey Professor at the Santa Fe Institute. In 1991, together with Norman Packard, he founded Prediction Company, a firm whose business is automatic trading of financial instruments using time series-based directional forecasting methods. He worked in the Theoretical Division and at the Center for Nonlinear Studies at Los Alamos National Laboratories, where he was Oppenheimer Fellow and founder of the Complex Systems Group. He received a BS from Stanford University and a PhD from the University of California at Santa Cruz, both in physics. Contact him at the Santa Fe Inst., 1399 Hyde Park Rd., Santa Fe, NM 87501.